# Perturbation theory for quantum-mechanical observables

J. D. Franson and Michelle M. Donegan
Johns Hopkins University
Applied Physics Laboratory
Laurel, MD 20723

*Abstract*:

The quantum-mechanical state vector is not directly observable even though it is the fundamental variable that appears in Schrodinger's equation. In conventional time-dependent perturbation theory, the state vector must be calculated before the experimentally-observable expectation values of relevant operators can be computed. We discuss an alternative form of time-dependent perturbation theory in which the observable expectation values are calculated directly and expressed in the form of nested commutators. This result is consistent with the fact that the commutation relations determine the properties of a quantum system, while the commutators often have a form that simplifies the calculation and avoids canceling terms. The usefulness of this method is illustrated using several problems of interest in quantum optics and quantum information processing.

# I. INTRODUCTION

One of the interesting features of quantum mechanics is the fact that the state vector is not directly observable even though it is the fundamental variable that appears in Schrodinger's equation. As a result, the state vector must be calculated first in conventional perturbation theory [1, 2], after which the experimentally-observable expectation values of the relevant operators can be evaluated. That approach seems conceptually unappealing from a computational point of view and it often complicates higher-order perturbation calculations, since it usually leads to a large number of canceling terms.

In this paper, we derive an alternative form of time-dependent perturbation theory in which the observable quantities are calculated directly without first calculating the perturbed state vector. The results obtained are expressed in terms of the commutators of the relevant operators, which is appealing in view of the fact that the commutation relations determine all of the properties of a quantum system. In addition, there are many problems of interest in which the commutators have properties that simplify the calculations and avoid the cancellation of higher-order terms.

The approach that we describe here is mathematically equivalent to a generalization of the Baker-Hausdorff theorem [3-5]. The relevant equations are derived using a limiting procedure in Section II, where their application to perturbation theory is described. Commutators are used in other forms of perturbation theory, such as the Magnus expansion [6] or Wick's theorem [7-9], but those approaches are not the same as that presented here, as we discuss in Section III. Although this approach to perturbation theory is straightforward and useful, it does not appear to be well known and we have only been able to find one reference to a similar procedure [10], which did not discuss any possible applications. The usefulness of this method is illustrated by applying it to several problems of interest in quantum optics and quantum information processing in Section IV. Our results are summarized in Section V.

## II. GENERALIZATION OF THE BAKER-HAUSDORFF THEOREM

In the Schrodinger picture, the time-evolution of the state vector $|\psi(t)>$ is given by Schrodinger's equation

$$\frac{d|\psi(t)>}{dt} \;=\; \frac{1}{i\hbar}\hat{H}(t)|\psi(t)> \tag{1}$$

where $\hat{H}(t)$ is the Hamiltonian. For the case where the Hamiltonian is independent of time, Eq. (1) has a formal solution given by

$$|\psi(t)> \;=\; e^{-i\hat{H}t/\hbar}|\psi_0> \tag{2}$$

Here the initial time has been taken to be 0 with $|\psi_0>$ the corresponding value of the state vector. The expectation value $<\psi(t)|\hat{O}|\psi(t)>$ of an operator $\hat{O}$ can then be evaluated using the Baker-Hausdorff theorem [3-5]

$$e^{\hat{A}}\hat{O}e^{-\hat{A}} = \hat{O}+[\hat{A},\hat{O}]+\frac{1}{2!}[\hat{A},[\hat{A},\hat{O}]]+\frac{1}{3!}[\hat{A},[\hat{A},[\hat{A},\hat{O}]]]+... \tag{3}$$

By replacing $\hat{A}$ with $i\hat{H}t/\hbar$ and taking the expectation value in state $|\psi(t)>$, we obtain

$$<\psi(t)|\hat{O}|\psi(t)> = <\psi_0|\hat{O}|\psi_0>+\frac{it}{\hbar}<\psi_o|[\hat{H},\hat{O}]|\psi_0>+\frac{1}{2!}\left(\frac{it}{\hbar}\right)^2<\psi_0|[\hat{H},[\hat{H},\hat{O}]]|\psi_0>+... \tag{4}$$

In most cases of interest, the operator $\hat{O}$ is independent of time in the Schrodinger picture, but Eq. (4) applies equally well if $\hat{O}$ is time-dependent.

As in conventional perturbation theory, it may be convenient to first make a unitary transformation to the interaction picture [2, 12] in order to ensure that the Hamiltonian is small in some sense, which typically increases the rate of convergence of the expansion of Eq. (4) . Schrodinger's equation and all of the results that follow still apply provided that the Hamiltonian in Eq. (1) is replaced by the interaction Hamiltonian and all operators are in the interaction picture (which is equivalent to Heisenberg operators in the absence of any interaction). For simplicity, we will either use the Schrodinger picture or assume that such a transformation has been performed but without changing the notation of the operators to denote the interaction picture. Operators in the Heisenberg picture will be denoted by a subscript H.

The perturbation expansion of Eq. (4) provides a convenient way to evaluate observables whenever the commutators have simple properties of various kinds, as is discussed in Section IV. But many problems of interest in atomic physics and quantum optics involve time-dependent Hamiltonians due to the application of external fields or laser pulses, for example. In that case the original Baker-Hausdorff theorem can no longer be applied, since the formal solution to Schrodinger's equation is then given [1, 2] by

$$|\psi(t)>=\hat{T}\exp\left(-i\int_0^t\hat{H}(t^{/})dt^{/}/\hbar\right)|\psi_0>  \tag{5}$$

where $\hat{T}$ is the time-ordering operator that re-orders the Hamiltonians in chronological order. One of the main results of this paper is that the Baker-Hausdorff theorem can be generalized to

$$\left( \hat{T} \exp(-i \int_0^t \hat{H}(t')dt'/\hbar) \right)^{-1} \hat{O} \left( \hat{T} \exp(-i \int_0^t \hat{H}(t'')dt''/\hbar) \right) = \tag{6}$$

$$\hat{O} + \frac{1}{i\hbar} \int_0^t dt' [\hat{O}, \hat{H}(t')] \; + \; \frac{1}{(i\hbar)^2} \int_0^t dt' \int_0^{t'} dt'' [[\hat{O}, \hat{H}(t')], \hat{H}(t'')] \; + \; ...$$

as will be shown below. As a result, the expectation value of an operator can be written in the case of a time-dependent Hamiltonian as

$$<\psi(t)|\hat{O}|\psi(t)> \; = \; <\psi_0|\hat{O}|\psi_0> + \frac{1}{i\hbar} \int_0^t dt' <\psi_0|[\hat{O}, \hat{H}(t')]|\psi_0> \tag{7}$$

$$+ \; \frac{1}{(i\hbar)^2} \int_0^t dt' \int_0^{t'} dt'' <\psi_0|[[\hat{O}, \hat{H}(t')], \hat{H}(t'')]|\psi_0> \; + \; ...$$

Equation (6) gives the Heisenberg operator $\hat{O}_H(t)$ as an expansion in commutators involving the Schrodinger operators $\hat{O}$ and $\hat{H}(t)$, while Eq. (7) gives the corresponding expectation value in the Schrodinger picture. It should be noted that Eqs. (6) and (7) do not contain any of the factorial terms that appear in Eqs. (3) and (4).

Eq. (7) is analogous to the usual perturbation formula [1, 2] for the state vector:

$$|\psi(t)> \; = \; |\psi_0> \; + \; \frac{1}{i\hbar} \int_0^t dt' \hat{H}(t')|\psi_0> \tag{8}$$

$$+ \; \frac{1}{(i\hbar)^2} \int_0^t dt' \int_0^{t'} dt'' \hat{H}(t')\hat{H}(t'')|\psi_0>$$

except that Eq. (7) involves commutators rather than the Hamiltonian itself. It seems somewhat surprising that this result does not appear in most texts on quantum mechanics. In particular, Eq. (7) explicitly shows that all observable properties of a system depend only on the commutation relations, as can be inferred in other ways.

One might expect that the most straightforward way to derive Eq. (7) would be to substitute Eq. (8) for the state vector into both the bra and ket of the desired expectation value:

$$<\psi(t)|\hat{O}|\psi(t)> = \left(|\psi_0>+\frac{1}{i\hbar}\int_0^t dt'\hat{H}(t')|\psi_0>+\frac{1}{(i\hbar)^2}\int_0^t dt'\int_0^{t'}dt''\hat{H}(t')\hat{H}(t'')|\psi_0>+...\right)^\dagger$$

$$\times\hat{O}\left(|\psi_0>+\frac{1}{i\hbar}\int_0^t dt'\hat{H}(t')|\psi_0>+\frac{1}{(i\hbar)^2}\int_0^t dt'\int_0^{t'}dt''\hat{H}(t')\hat{H}(t'')|\psi_0>+...\right)$$

(9)

But multiplying out all the terms and collecting those of a given order produces integrals whose limits do not correspond to those of Eq. (7). For example, in second order we have the second order term in the bra multiplied by the zero order term in the ket, plus the first order term in the bra multiplied by the first order term in the ket, and so forth. The product of the two first-order terms corresponds to

$$\int_0^t dt'\int_0^t dt'' = \int_0^t dt'\left(\int_0^{t'}dt''+\int_{t'}^t dt''\right) = \int_0^t dt'\int_0^{t'}dt'' + \int_0^t dt''\int_0^{t''}dt'$$

(10)

in an obvious simplified notation, where the limits of the integrals have been manipulated to give the same form as those in Eq. (7). We have verified Eq. (7) up to third order using this method, which becomes increasingly tedious for higher orders. To our knowledge, no proof by induction has been given using this method.

Our proof of Eqs. (6) and (7) is based instead on a limiting process [13] in which the time interval from 0 to t is divided into N equal intervals of length $\Delta t=t/N$. In the limit of large N, $\hat{H}(t)$ will be nearly constant over any given time interval $t_i$, where $i$ is an index with values 1 to $N$. As a result, the expectation value of $\hat{O}$ can be found by propagating over each of the successive intervals using Eq. (2):

$$<\psi(t)|\hat{O}|\psi(t)> = <\psi_0|e^{i\hat{H}(t_1)\Delta t/\hbar}...e^{i\hat{H}(t_n)\Delta t/\hbar}$$
$$\times\hat{O}\,e^{-i\hat{H}(t_n)\Delta t/\hbar}...e^{-i\hat{H}(t_1)\Delta t/\hbar}|\psi_0>$$

(11)

The middle three terms in the above expression can be rewritten to first order in $\Delta t$ as

$$e^{i\hat{H}(t_n)\Delta t/\hbar}\,\hat{O}\,e^{-i\hat{H}(t_n)\Delta t/\hbar} = \hat{O}+[\hat{O},\hat{H}(t_n)\Delta t/i\hbar]$$
$$\equiv \hat{O}+[\hat{O},\hat{H}_n]$$

(12)

where we have simplified the notation by defining $\hat{H}_n=\hat{H}(t_n)\Delta t/i\hbar$. This result can be obtained by first-order expansion of the exponentials or from the first term of the original Baker-Hausdorff theorem. Inserting Eq. (12) into Eq. (11) reduces the expectation value to

$$
\begin{aligned}
<\psi(t)|\hat{O}|\psi(t)> \ &= \ <\psi_0|e^{i\hat{H}(t_1)\Delta t/\hbar}...e^{i\hat{H}(t_{n-1})\Delta t/\hbar} \\
&\times (\hat{O}+[\hat{O},\hat{H}_n])\, e^{-i\hat{H}(t_{n-1})\Delta t/\hbar}...e^{-i\hat{H}(t_1)\Delta t/\hbar}|\psi_0>
\end{aligned}
\tag{13}
$$

This process can then be repeated for the middle three terms of Eq. (13):

$$
\begin{aligned}
e^{i\hat{H}(t_{n-1})\Delta t/\hbar}(\hat{O}+[\hat{O},\hat{H}_n])\, e^{-i\hat{H}(t_{n-1})\Delta t/\hbar} \ &= \ \hat{O}+[\hat{O},\hat{H}_n]+ \\
&[\hat{O},\hat{H}_{n-1}]+[[\hat{O},\hat{H}_n],\hat{H}_{n-1}]
\end{aligned}
\tag{14}
$$

where the double commutator is obtained when the original commutator $[\hat{O},\hat{H}_n]$ from Eq. (13) is commuted with $\hat{H}_{n-1}$ in the expansion of the next exponential. Repeating this process $N$ times gives a typical term of order $m$ in which the commutator was taken at m time intervals i, j, ... , k:

$$
<\psi(t)|\hat{O}|\psi(t)>^{(m)}_{ij...k} \ = \ <\psi_0|[[[\hat{O},\hat{H}_i],\hat{H}_j],...,\hat{H}_k]|\psi_0>
\tag{15}
$$

Summing these terms over all of the possible times at which the commutators can be taken gives

$$
<\psi(t)|\hat{O}|\psi(t)>^{(m)} \ = \ \sum_{i=1}^{N}\sum_{j=1}^{i}...\sum_{k=1}^{l}<\psi_0|[[[\hat{O},\hat{H}_i],\hat{H}_j]...,\hat{H}_k]|\psi_0>
\tag{16}
$$

Taking the limit of large $N$ converts the sums to integrals and gives the desired result:

$$
\begin{aligned}
&<\psi(t)|\hat{O}|\psi(t)>^{(m)}= \\
\frac{1}{(i\hbar)^m}\int_0^t dt'\int_0^{t'}dt''...&\int_0^{t'''}dt''''<\psi_0|[[[\hat{O},\hat{H}(t')],\hat{H}(t'')],...,\hat{H}(t'''')]|\psi_0>
\end{aligned}
\tag{17}
$$

which is equivalent to Eq. (7). It should be noted that the first commutator is taken with the Hamiltonian evaluated at the most recent time $t'$.

Although the above derivation included expectation values, it should be apparent that the same procedure could be used without taking expectation values to obtain the expansion of the Heisenberg operator $\hat{O}_H(t)$ shown in Eq. (6). The equivalence of the time-ordered products with successive propagation over small time intervals [13] follows from Dyson's results [1].

In his text on dispersion relations, Nishijima [10] independently derived a similar result for quantum field operators using a different method; no references to earlier work were included and there was no discussion of potential applications. This is the only reference to related work that we have been able to find, although we suspect that there may be others given the straightforward nature of these results.

## III.  COMPARISON WITH OTHER METHODS

One might suspect that these results could be derived more simply using other methods, such as Taylor series expansion or iteration. In this section, we show that those methods do not produce the same results as were derived above. Other perturbation-theory approaches that involve the use of commutators will also be compared to the method described here.

In the Heisenberg picture, the time-evolution of an operator $\hat{O}_H(t)$ is given as usual [2] by

$$\frac{d\hat{O}_H(t)}{dt} = \frac{1}{i\hbar}[\hat{O}_H(t), \hat{H}_H(t)] + \frac{\partial \hat{O}_H(t)}{\partial t} \qquad (18)$$

where $\hat{O}_H(t)$ is defined as

$$\hat{O}_H(t) = \left(\hat{T}\exp(-i\int_0^t \hat{H}(t')dt'/\hbar)\right)^{-1} \hat{O}\left(\hat{T}\exp(-i\int_0^t \hat{H}(t'')dt''/\hbar)\right) \qquad (19)$$

and the transition operator $\hat{U}(t,0)$ is given by

$$\hat{U}(t,0) = \hat{T}\exp(-i\int_0^t \hat{H}(t')dt'/\hbar) \qquad (20)$$

It should be noted that it is the Hamiltonian $\hat{H}_H(t)$ in the Heisenberg picture that appears in Eq. (18), as can be shown by differentiating Eq. (19) and inserting $\hat{U}\hat{U}^{-1}=\hat{I}$ between the two Schrodinger operators, where $\hat{I}$ is the identity operator . For a time-dependent Hamiltonian, $\hat{H}_H(t) \neq \hat{H}(t)$, which will have important implications in the discussion that follows.

We first consider the possibility of combining the operator equation of motion of Eq. (18) with a Taylor series expansion to derive an expression for $\hat{O}_H(t)$. The difficulty with that approach is the last term in Eq. (18), which includes the effects of any explicit time-dependence of the operator being differentiated. Even if the original operator $\hat{O}$ of interest does not have any time dependence in the Schrodinger picture, the second and higher derivatives will involve the time dependence of the Hamiltonian $\hat{H}(t)$:

$$\frac{d^2\hat{O}_H}{dt^2} = \frac{d}{dt}\left(\frac{1}{i\hbar}[\hat{O}_H, \hat{H}_H(t)]\right)$$

$$= \frac{1}{(i\hbar)^2}[[\hat{O}_H, \hat{H}_H(t)], \hat{H}_H(t)] + \frac{1}{i\hbar}\frac{\partial[\hat{O}_H, \hat{H}_H(t)]}{\partial t} \tag{21}$$

It can be seen that a Taylor series expansion would include terms other than the nested commutators of Eq. (6) and that a Taylor series expansion is not equivalent to the results presented above.

Since conventional time-dependent perturbation theory [1, 2] is based on an iterative solution to Schrodinger's equation, one might suspect that our results could be derived from an iterative solution to the operator equation of motion, Eq. (18). Here we will assume that the operator $\hat{O}$ of interest has no time-dependence in the Schrodinger equation, so that the partial derivative is zero in this case. Assuming that $\hat{H}_H(t)$ is small in some sense, the first-order solution can be obtained as usual by inserting the zero-order value of $\hat{O}_H(t)$ into the right-hand side of Eq. (18) and integrating:

$$\hat{O}_H(t) = \hat{O} + \frac{1}{i\hbar}\int_0^t dt'[\hat{O}, \hat{H}_H(t')] \tag{22}$$

The second-order term can then be obtained by inserting this result into Eq. (18) and integrating again:

$$\hat{O}_H(t) = \hat{O} + \frac{1}{i\hbar}\int_0^t dt'[\hat{O}, \hat{H}_H(t')] + \frac{1}{(i\hbar)^2}\int_0^t dt'\int_0^{t'} dt''[[\hat{O}, \hat{H}_H(t'')], \hat{H}_H(t')] + ... \tag{23}$$

This process can be repeated to give an expression for $\hat{O}_H(t)$ to any desired order in $\hat{H}_H(t)$. The order of the operators in the commutator of Eq. (18) can be reversed with the addition of a minus sign, so that Eq. (23) can also be written as

$$\hat{O}_H(t) = \hat{O} + \frac{1}{(-i\hbar)}\int_0^t dt'[\hat{H}_H(t'), \hat{O}] + \frac{1}{(-i\hbar)^2}\int_0^t dt'\int_0^{t'} dt''[\hat{H}_H(t'), [\hat{H}_H(t''), \hat{O}]] + ... \tag{24}$$

Eq. (24) has been previously derived [14, 15] for the special case in which $\hat{O}$ is taken to be the density operator.

At first glance, Eqs. (23) and (24) appear to be very similar to Eq. (6) in the preceding section, since they both involve integrals of nested commutators. But the first commutator is taken with the Hamiltonian evaluated at the earliest time $t''$ in Eq. (24), whereas the first commutator involves the Hamiltonian evaluated at the most recent time in our results of Eqs (6) and (7). As a result, the commutators are totally different in the two cases, as can be seen if we let operators $\hat{A}$, $\hat{B}$, and $\hat{C}$ denote the Hamiltonian evaluated at three different times, with $\hat{A}$ the most recent. Our original result in Eq. (6) then corresponds to commutators of the form

$$[[[\hat{O}, \hat{A}], \hat{B}], \hat{C}] \;=\; \begin{aligned}&\hat{O}\hat{A}\hat{B}\hat{C} - \hat{A}\hat{O}\hat{B}\hat{C} - \hat{B}\hat{O}\hat{A}\hat{C} + \hat{B}\hat{A}\hat{O}\hat{C}\\ &-\hat{C}\hat{O}\hat{A}\hat{B} + \hat{C}\hat{A}\hat{O}\hat{B} + \hat{C}\hat{B}\hat{O}\hat{A} - \hat{C}\hat{B}\hat{A}\hat{O}\end{aligned} \qquad (25)$$

while Eq. (24) corresponds to

$$[\hat{A}, [\hat{B}, [\hat{C}, \hat{O}]]] \;=\; \begin{aligned}&\hat{A}\hat{B}\hat{C}\hat{O} - \hat{A}\hat{B}\hat{O}\hat{C} - \hat{A}\hat{C}\hat{O}B + \hat{A}\hat{O}\hat{C}\hat{B}\\ &-\hat{B}\hat{C}\hat{O}\hat{A} + \hat{B}\hat{O}\hat{C}\hat{A} + \hat{C}\hat{O}\hat{B}\hat{A} - \hat{O}\hat{C}\hat{B}\hat{A}\end{aligned} \qquad (26)$$

It can be seen that none of the terms in Eq. (25) are the same as any of the terms of Eq. (26), which shows that the commutator expressions are completely different in the two cases. More importantly, the results derived by iteration in Eqs. (23) and (24) give an expansion of $\hat{O}_H(t)$ in powers of $\hat{H}_H(t)$, whereas our original result in Eq. (6) gives an expansion in powers of $\hat{H}(t)$. In most situations of interest in atomic physics or quantum optics, the Hamiltonian has a simple form in the Schrödinger picture, so that $\hat{H}(t)$ is the starting point for most calculations. Unless the problem is sufficiently simple that an analytic form can be found for $\hat{H}_H(t)$, the calculation of the Hamiltonian in the Heisenberg picture may be as difficult as the calculation of $\hat{O}_H(t)$ itself. In fact, it may be necessary to calculate $\hat{H}_H(t)$ using the perturbation technique of Eq. (6) before Eqs. (23) or (24) could be applied. As a result, our results in the preceding section appear to be more useful for actual calculations than are those derived using iteration.

Commutators also appear in other forms of time-dependent perturbation theory, including the Magnus expansion [6] and Wick's theorem [7]. Those approaches are widely used in scattering theory or quantum field theory, where the primary interest is in the scattering matrix $\hat{S} = \hat{U}(+\infty, -\infty)$. The transition operator obeys the differential equation

$$\frac{d\hat{U}(t,0)}{dt} \;=\; \frac{1}{i\hbar}\hat{H}(t)\hat{U}(t,0) \qquad (27)$$

whose form is completely different from that for observable operators (Eq. 18). Iteration of Eq. (27) leads to Dyson's expansion [1] and to Eq. (20), which do not involve commutators. In quantum field theory, Wick's theorem [7] is commonly used to reduce the time-ordered operators to normal order, which results in the familiar Feynman diagrams [8, 9, 11]. The contraction of two operators that is used in Wick's theorem involves commutators but it does not result in nested commutators of the kind that appear in our results.

Other forms of perturbation theory obtain exponential solutions [6, 16-19] for the transition operator of the form

$$\hat{U}(t,0) = e^{\hat{\Omega}(t)} \tag{28}$$

Magnus [6] has given a method for obtaining a perturbative expansion for $\hat{\Omega}(t)$ that also involves nested commutators. That approach has been applied to a large number of different problems, since equations with the same general form as Eq. (27) appear in a variety of classical and quantum systems. The Magnus expansion does not have the same form as our result, however, since it corresponds to a solution for the transition matrix rather than for observable operators. The calculation of the coefficients in the Magnus expansion are related to the Campbell-Baker-Hausdorff theorem [20-22] and are more complicated than our expansion, which is related to the simpler Baker-Hausdorff theorem.

Bialynicki-Birula, Mielnik, and Plebanski [23] have given a formal solution for any function of the transition operator. When applied to the logarithm of the transition operator, this method gives an expansion for $\hat{\Omega}(t)$ that is related to the Magnus expansion and also involves nested commutators. They have also applied this approach to find a closed form for the phase operator, for example. Their method is more general than ours and could presumably be used to provide an alternative derivation of Eq. (6), although that has not been done to the best of our knowledge.

Commutators also play a role in various forms of steady-state perturbation theory [24-26], which often involve the Campbell-Baker-Hausdorff theorem. Those approaches are unrelated to ours, which is concerned instead with time-dependent perturbation theory.

## IV. EXAMPLES OF APPLICATIONS IN QUANTUM OPTICS

We independently derived Eq. (7) a number of years ago and have used it on several problems of interest since then. An approach of this kind may be useful whenever the operators have some simple properties that allow the nested commutators to be readily evaluated. The simplest case is when the commutators are all zero under some condition, which occurs, for example, in proofs of causality. That was the original motivation for our work in this area, as will be discussed in a separate paper. Another situation where the method is useful is when the commutators themselves are nonzero but the second-order commutators vanish:

$$[[\hat{O}, \hat{H}(t')], \hat{H}(t'')] = 0 \tag{29}$$

The expansion of Eqs. (6) and (7) will then include only the first-order term. More generally, it may be that the commutators obey a recursion relation of some kind in which the higher-order commutators can all be related to some combination of lower-order commutators. All of these situations occur in quantum optics, as we now illustrate with two examples.

Our first example involves the possibility of stimulated emission by a classical current distribution $\vec{j}_c(\vec{r},t)$ that is subjected to an intense external electric field $\vec{E}_{ext}(\vec{r},t)$, as illustrated in Figure 1. It will be convenient to assume that $\vec{E}_{ext}(\vec{r},t)$ is generated by another classical current distribution $j_{ext}(\vec{r},t)$. It is well known that a strong external field can increase the rate of emission of photons via stimulated emission, since the matrix elements for emission increase when there are large numbers of photons initially present. In particular,

$$\hat{a}^{\dagger}_{\vec{k}}\,|n_{\vec{k}}> \;=\; \sqrt{n_{\vec{k}}+1}\,|n_{\vec{k}}+1> \tag{30}$$

where $\hat{a}^{\dagger}_{\vec{k}}$ creates a photon with wave vector $\vec{k}$ and $|n_{\vec{k}}>$ represents a state with $n_{\vec{k}}$ photons in that mode. The question is whether or not the presence of $\vec{E}_{ext}(\vec{r},t)$ will increase the amount of electromagnetic radiation emitted by $\vec{j}_c(\vec{r},t)$, since $n_{\vec{k}}$ and the relevant matrix elements will both be very large. Presumably that cannot occur since Maxwell's equations are linear, but we will use this example to illustrate the methods described above.

The interaction Hamiltonian for this system is given [2] by

$$\hat{H}(t) \;=\; -\frac{e}{c}\int \vec{j}(\vec{r},t)\cdot\hat{A}(\vec{r},t)d^3\vec{r} \;\;=\; \int \vec{j}(\vec{r},t)\cdot\sum_{\vec{k}}\alpha(\vec{k})\vec{\lambda}_k(\hat{a}e^{\,i(\vec{k}\cdot\vec{r}-\omega t)}+\hat{a}^{\dagger}e^{\,-i(\vec{k}\cdot\vec{r}-\omega t)})\,d^3\vec{r} \tag{31}$$

Here $\hat{A}(\vec{r},t)$ is the vector potential operator, $\vec{j}(\vec{r},t)=\vec{j}_c(\vec{r},t)+\vec{j}_{ext}(\vec{r},t)$ is the total current, e is the charge of the electron, c the speed of light, $\vec{\lambda}_{\vec{k}}$ represents two orthogonal polarization modes, $\omega=ck$, and all of the constants have been included in the coefficients $\alpha(\vec{k})$ which are of no interest here. Our goal is to calculate the expectation value of the total electric field at a test point $\vec{r}$. From Eq. (7), the expectation value of the electric field operator $\hat{E}(\vec{r},t)$ is given by

$$\begin{aligned}
<\hat{E}(\vec{r},t)> \;=\; &<0|\hat{E}(\vec{r},t)|0> \;+\; \frac{1}{(i\hbar)}\int_0^t dt'<0|[\hat{E}(\vec{r},t),\hat{H}(t')]|0> \\
&+\; \frac{1}{(i\hbar)^2}\int_0^t dt'\int_0^{t'} dt''<0|[[\hat{E}(\vec{r},t),\hat{H}(t')],\hat{H}(t'')]|0> \;+\; ...
\end{aligned} \tag{32}$$

where we have simplified the notation by writing $|0>=|\psi_0>$. In this case the commutator is a complex number and not an operator, since $\hat{E}(\vec{r},t)$ and $\hat{H}(t)$ are both linear in $\hat{a}$ and $\hat{a}^{\dagger}$:

$$[\hat{E}(\vec{r},t),\hat{H}(t')] \;=\; \sum_{\vec{k},\vec{k}'}\alpha'(\vec{k},\vec{k}')[\hat{a}^{\dagger}_{\vec{k}},\hat{a}_{\vec{k}'}] \;=\; \sum_{\vec{k},\vec{k}'}\alpha'(\vec{k},\vec{k}')\delta(\vec{k}-\vec{k}') \tag{33}$$

where $\alpha'(\vec{k},\vec{k}')$ is another set of coefficients. The second and all higher commutators are zero as a consequence and only the first term in Eq. (32) can contribute. That term is linear in $\hat{H}(t)$, which can be written as the sum of the contributions from the two current distributions $\vec{j}_c$ and $\vec{j}_{ext}$:

$$\hat{H}(t) \ = \ \hat{H}_c(t) + \hat{H}_{ext}(t) \tag{34}$$

As a result, the expectation value of the electric field is the sum of the fields from the two current distributions calculated in the absence of each other

$$
\begin{aligned}
<\hat{E}(\vec{r},t)> \ = \ &<0|\hat{E}(\vec{r},t)|0> \ + \ \frac{1}{(i\hbar)}\int_0^t dt'<0|[\hat{E}(\vec{r},t),\hat{H}_c(t')]|0> \\
&+ \ \frac{1}{(i\hbar)}\int_0^t dt'<0|[\hat{E}(\vec{r},t),\hat{H}_{ext}(t')]|0>
\end{aligned}
\tag{35}
$$

The presence of $\vec{E}_{ext}(\vec{r},t)$ has no effect on the field emitted by $\vec{j}_c$, despite the large number of photons initially present. Roughly speaking, the increased magnitude of the matrix elements does increase the rate of emission, but the absorption of radiation is increased by a correspondingly large amount. This result is exact and well known, since it can be obtained in several other ways, but it does illustrate the way in which these methods can be applied.

As a second example, we consider the question of whether or not the action of a set of beam splitters can be nonlinear when the input fields are non-classical and contain only a few photons. As illustrated in Figure 2, several non-classical beams of light are assumed to be incident on some combination of beam splitters. Two of the incident beams have been labeled input 1 and input 2, with two output beams labeled output 1 and output 2 for reasons that will become apparent shortly. In classical physics, ideal beam splitters are linear devices, since the two output fields from a given beam splitter are linear combinations of the two input fields. But when the problem is treated quantum-mechanically, the matrix elements will involve factors of $\sqrt{n_i+1}$, for example, where $n_i$ is the number of photons in mode $i$. Since this is a nonlinear function of $n_i$, it may not be entirely apparent whether or not the system is linear at the single-photon level [27, 28].

The Hamiltonian for this system can be written in a way that makes it equivalent to that of a set of coupled harmonic oscillators:

$$\hat{H}(t) \ = \ \sum_i (\hat{a}_i^\dagger \hat{a}_i + 1/2)\hbar\omega_i(t) \ + \ \sum_{i \neq j} \epsilon_{ij}(t)(\hat{a}_i^\dagger \hat{a}_j + \hat{a}_j^\dagger \hat{a}_i) \tag{36}$$

where the coupling between modes $i$ and $j$ is proportional to the constant $\epsilon_{ij}(t)$. For the sake of discussion, we have assumed that the coupling constants $\epsilon_{ij}(t)$ and the frequencies $\omega_i(t)$ are all time-dependent, in which case an analytic solution may not be available [29]. (Analogous Hamiltonians can be implemented in an atomic medium using laser pulses [27, 28] ). We will focus on the question of whether or not the number $n_2$ of photons in output port 2 can depend nonlinearly on the number of photons $n_{10}$ and $n_{20}$ initially present in inputs 1 and 2. Questions of this kind are of practical importance in quantum computing and other forms of quantum information processing.

From Eq. (7), the expectation value of $n_2$ is given by

$$\langle \hat{n}_2 \rangle = \langle 0 | \hat{n}_2 | 0 \rangle + \frac{1}{(i\hbar)} \int_0^t dt' \langle 0 | [\hat{n}_2, \hat{H}(t')] | 0 \rangle$$
$$+ \frac{1}{(i\hbar)^2} \int_0^t dt' \int_0^{t'} dt'' \langle 0 | [[\hat{n}_2, \hat{H}(t')], \hat{H}(t'')] | 0 \rangle + \ldots$$

(37)

In evaluating the first-order commutator, the product of the two original operators involves four factors of $\hat{a}_i$ or $\hat{a}_i^\dagger$. But two factors of $\hat{a}_i$ or $\hat{a}_i^\dagger$ are eliminated using the commutation relations, so that the commutator itself is left with only two operators:

$$[\hat{n}_2, \hat{H}] = [\hat{a}_2^\dagger \hat{a}_2, \sum_{ij} \alpha_{ij} \hat{a}_i^\dagger \hat{a}_j] = \sum_{ij} \alpha_{ij}' \hat{a}_i^\dagger \hat{a}_j$$

(38)

where $\alpha_{ij}$ and $\alpha_{ij}'$ represent the corresponding coefficients, which are of no interest here. This pattern is repeated when calculating the second and higher-order commutators. For example,

$$[[\hat{n}_2, \hat{H}(t')], H(t'')] = [\sum_{i'j'} \alpha_{i'j'}' \hat{a}_{i'}^\dagger \hat{a}_{i'}, \sum_{ij} \alpha_{ij} \hat{a}_i^\dagger \hat{a}_j] = \sum_{i'j'} \alpha_{i'j'}'' \hat{a}_{i'}^\dagger \hat{a}_{j'}$$

(39)

As a result, all of the terms in the expansion of $\langle \hat{n}_2 \rangle$ reduce to a bilinear combination of the creation and annihilation operators:

$$\langle \psi(t) | \hat{n}_2 | \psi(t) \rangle = \sum_{ij} \beta_{ij} \langle \psi_0 | \hat{a}_i^\dagger \hat{a}_j | \psi_0 \rangle$$

(40)

where the coefficients $\beta_{ij}$ are related to the $\alpha_{ij}$. If all the input beams contain photon number states, which is the case of most interest in quantum computing applications, then $\langle \psi_0 | \hat{a}_i^\dagger \hat{a}_j | \psi_0 \rangle = 0$ unless $i=j$ and the final expectation value is a linear combination of the initial expectation values:

$$\langle\psi(t)|\hat{n}_2|\psi(t)\rangle \;=\; \sum_i \beta_{ii}\langle\psi_0|\hat{n}_i|\psi_0\rangle \tag{41}$$

Equation (41) shows that the expectation values of the outputs are linear functions of the inputs, just as is the case classically.

It is tempting to conclude from this that beam splitters are linear elements after all, even when dealing with single photons.  But expectation values give only the average result of a measurement and several groups [30-33] have recently shown that linear optical elements can be used to perform quantum logic operations provided that the events are post-selected based on the results of measurements.  For example, we have recently shown [32, 33] that a probabilistic controlled-NOT gate can be implemented using polarizing beam splitters, additional photons (ancilla), and single-photon detectors that measure the ancilla output modes similar to those shown in Figure 2.  Post-selection measurements of this kind can be represented by more complex operators such as $\hat{n}_2\hat{D}_1\hat{D}_2\hat{D}_3\hat{D}_4$, where $\hat{D}_i = 0$ or 1 depending on the output of four detectors labeled with the index $i$.  The expectation values of more complex operators of this kind can also be evaluated using the methods described above, which is one of the reasons for our interest in this approach.

## V.  SUMMARY AND CONCLUSIONS

We have presented a perturbative method for calculating the expectation value of quantum-mechanical operators for the case in which the Hamiltonian is a function of time.  The perturbed state vector need not be calculated and the results are given directly in the form of a series of integrals of nested commutators.  This provides a direct connection between the observable results of an experiment and the commutation relations, which determine the properties of a quantum-mechanical system.  As a practical matter, the approach is often useful whenever the commutators have sufficiently simple properties.  The cancellation of higher-order terms that often occurs in standard time-dependent perturbation theory can frequently be avoided.  Several examples of how these techniques may be applied to problems of interest in quantum optics or quantum information processing were presented.

Although commutators are used in other forms of perturbation theory, the approach presented here is not equivalent to what would be obtained from commonly-used techniques such as Taylor-series expansion or iteration, and we have only been able to find one related reference [10].  In any event, this approach does not appear to be widely used in atomic physics or quantum optics, and we hope that this paper will serve to illustrate its potential usefulness.

We would like to thank B. C. Jacobs, S. P. Kim, T. B. Pittman, and W. P.  Schleich for their comments.  This work was supported by the Office of Naval Research.

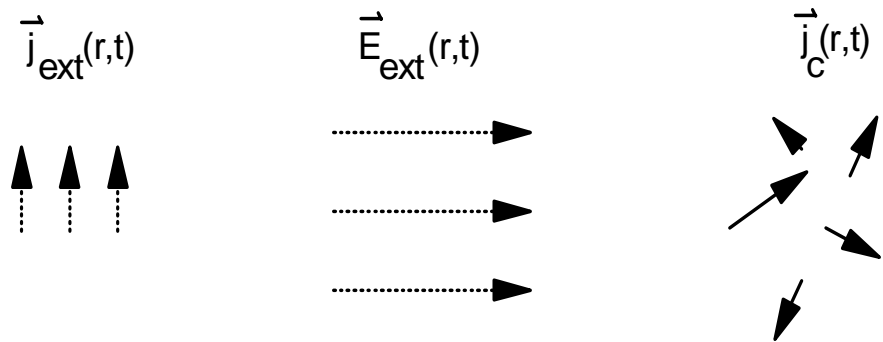

Fig. 1  An example of a simple situation in quantum optics where the perturbation theory
described here can be easily applied.  A classical current distribution $\vec{j}_c(\vec{r},t)$ is subjected to
an intense external electric field $\vec{E}_{ext}(\vec{r},t)$ that in turn is generated by another current
distribution $\vec{j}_{ext}(\vec{r},t)$.  The potential role of stimulated emission in the radiation of photons
by $\vec{j}_c(\vec{r},t)$ is to be investigated.  In this case, the second and higher-order commutators are
zero, so that the first term in Eq. (7) gives the exact result, which is linear and equivalent
to the classical field.

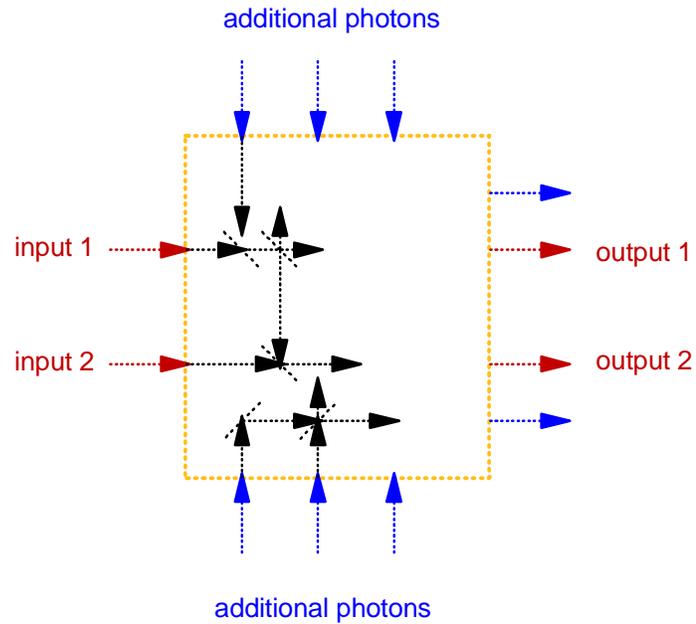

Fig. 2. An example of a system in which the higher-order commutators can be related in a simple way to the first-order commutators. Single photons enter a device containing a large number of linear elements, such as beam splitters, where they are combined with a number of additional photons (ancilla) to produce two outputs. Eq. (7) can be used to show that the expectation value of the output of this system is a linear function of the inputs, despite the fact that the matrix elements depend nonlinearly on the number of photons. Devices of this kind can, however, be used in conjunction with post-selection to perform quantum logic operations such as a controlled-NOT, which are inherently nonlinear operations.